\documentclass[fleqn,10pt]{olplainarticle}
\usepackage{appendix}
\usepackage{siunitx}
\usepackage{float}

\title{A finite difference scheme for integrating the Takagi-Taupin equations on an arbitrary orthogonal grid}

\author[1]{Mads Carlsen}
\author[1]{Hugh Simons}

\affil[1]{Department of Physics, Technical University of Denmark, 2800 Kgs. Lyngby, Denmark}


\begin{abstract}
Calculating dynamical diffraction patterns for X-ray topography and similar x-ray scattering-imaging techniques require the numerical integration of the Takagi-Taupin equations. This is usually performed with a simple second order finite difference scheme on a sheared computational grid with two of the axes aligned with the wave vectors of the incident and scattered beams respectively. This dictates, especially at low scattering angles, an oblique grid of uneven step-sizes. Here we present a finite difference scheme that carries out this integration in slab-shaped samples on an arbitrary orthogonal grid by implicitly utilizing Fourier interpolation. The scheme achieves the expected second order convergence and a similar error to the traditional approach on similarly dense grids.
\end{abstract}

\begin{document}

\flushbottom
\maketitle
\thispagestyle{empty}

\section*{Introduction}

In x-ray diffraction imaging, simulations based on the propagation of coherent wave fronts are often used to evaluate the viability of proposed techniques\cite{Pedersen2018}, and to investigate the effect of various experimental errors.\cite{Shabalin2017,Carnis2019,Holstad2022} In particular, with the new highly coherent x-ray sources (e.g. fourth-generation synchrotrons and free electron lasers), methods based on coherent wave-fronts are becoming increasingly useful.

When investigating large and near-perfect crystals, multiple scattering effects in the sample (dynamical diffraction) becomes important. Typically in diffraction imaging techniques, one tries to avoid dynamical effects (even if occasionally the dynamical effect are he subject of interest\cite{Rodriguez-Fernandez2021}) by using highly deformed samples, small grains or relying of the "weak beam approximation" i.e. measuring at the tails of the rocking curve.\cite{Shabalin2017, Holstad2022} In many cases, however, dynamical effects are unavoidable and therefore must be able to be simulated as well. 

The equations for propagating coherent wave fronts through deformed crystals are the Takagi-Taupin Equations (TTEs): a set of coupled 1st order PDEs that, in general, must be integrated numerically. When numerically integrating the TTEs (in the two beam case), it is natural to choose a computational grid where two of the axes are aligned with the wave-vector of the incident and scattered waves respectively. The use of this sheared grid may complicate matters by requiring the use of an connecting interpolation step when the scattering calculation is combined with other numerical methods used to generate the input for the crystal structure and the incident radiation, and when the calculated diffraction patterns are further input into simulations of the down-stream optics. This intermediate interpolation step can be both computationally expensive and may introduce artifacts.

For kinematical calculations, the authors of \cite{Li:to5204} have presented a method for carrying out scattering calculations using an orthogonal grid. This method implicitly use Fourier interpolation to avoid making cumulative interpolation errors, that would otherwise cause such a calculation to fail. Inspired by this, we present a method of integrating the dynamical Takagi-Taupin equations on an orthogonal grid, that also utilizes Fourier interpolation implicitly. 

%The most obvious way to do this is to interpolate the values of the crystal scattering function from the orthogonal grid, on which it is known, onto a sheared grid and then integrate the TTEs using established finite difference schemes. In order to avoid interpolation artifacts, which would be highly problematic for the simulation of down-stream optics, we would have to use Fourier interpolation. It is not clear how this extra interpolation step would impact the convergence behaviour of the established methods so rather we looked for a  systematic way to include this interpolation in the finite difference scheme itself. 
In the following, we present a way to do this based on exponential Rosenbock methods. The end result is similar to the mixed real-space/reciprocal-space methods called "multistep methods" used to model a wide range of optical problems. The method we arrive at will only be applicable for slab-like samples (two parallel surfaces and infinite extend in the orthogonal directions) in Laue geometry. 

\section{Dynamical diffraction}

The most general framework for treating dynamical diffraction from strained crystals is the Takagi-Taupin equations (TTE)[\cite{Takagi:a03704, doi:10.1143/JPSJ.26.1239,Taupin:a05601}] which, in the simplest two-beam case assuming $\sigma$ polarization and exact satisfaction of the Bragg condition, can be written as: \\
\begin{equation}
\begin{split}
2i(\mathbf{k}_0\cdot\nabla) E_0(\mathbf{r}) & = k^2(\chi_0 E_0(\mathbf{r}) + \chi_{\overline{h}}'(\mathbf{r})E_h(\mathbf{r}) )\\
2i(\mathbf{k}_h\cdot\nabla) E_h(\mathbf{r}) & = k^2(\chi_0 E_h(\mathbf{r}) + \chi_{h}'(\mathbf{r})E_0(\mathbf{r}))
\end{split}
\label{eq:TT}
\end{equation}

where $E_0$ and $E_h$  are the complex envelopes of the monochromatic fields of the incident and scattered beams respectively. $\mathbf{k}_0$ is the wave vector of the incident wave in vacuum and $\mathbf{k}_h = \mathbf{k}_0 + \mathbf{Q}$ is the wave-vector of the scattered beam. The choice of $\mathbf{k}_0$ when writing up the TTEs is arbitrary and leads to different versions of the TTEs. The other common choice is to choose $\mathbf{k}_0$ to be the wave vector of the refracted wave inside the crystal. \\

$\chi_h$ and $\chi_{\overline{h}}'$ are the spatially varying Fourier components of the electric susceptibility corresponding to the Bragg reflection with scattering vector $\mathbf{Q}$ and $\mathbf{-Q}$ respectively. They are related to the Fourier components of the perfect lattice of the undeformed crystal, $\chi_h$ and $\chi_{\overline{h}}$ through:

\begin{equation}
\begin{split}
\chi_h'(\mathbf{r}) &= \exp\big(i\mathbf{Q}\cdot\mathbf{u}(\mathbf{r})\big)\chi_{h} \\
\chi_{\overline{h}}'(\mathbf{r}) &= \exp\big(-i\mathbf{Q}\cdot\mathbf{u}(\mathbf{r})\big)\chi_{\overline{h}}
\end{split}
\end{equation}

where $\mathbf{u}(\mathbf{r})$ is the displacement field of the crystal. These constants are the macroscopic equivalents of the more often used form factors and are related to these through:\\

\begin{equation}
\chi_{h} = -\left( \frac{4\pi r_0}{k^2V_{\mathrm{u.c.}}} \right) F_h
\end{equation}

If we ignore the scattering terms, the equations \eqref{eq:TT} are a pair of convection equations and the solution involves the interpolation of the initial condition through the integration volume. Direct application of a finite-difference scheme in a Cartesian coordinate system would lead to interpolation errors accumulating at each step giving an unwanted dispersion of the initial condition. The traditional approach is therefore to solve the equation in an oblique coordinate system with the axes aligned with the incident and scattered wave-vectors. \\

The TTEs have been solved by finite difference integration on a structured grid of constant [\cite{Authier:a05857}] or varying step-sizes[\cite{Epelboin:a19607}] or by an iterative approach.[\cite{Bremer:a23062,PhysRevB.89.014104}] For certain symmetric geometries the sheared coordinate system coincides with a rectangular one.[\cite{PunegovRectangular, gup-82}] Recent efforts have also been made to calculate dynamical diffraction patterns from unstructured grids using a finite element approach. [\cite{Honkanen:te5029}] \\

% For hard x-rays ($E \geq 15 \mathrm{keV}$), the scattering angle can become small and the oblique coordinate system becomes inconvenient due to the small angle between the axes, which dictates a much shorter step size in the transverse directions that in the direction of propagation. Furthermore, numerical simulations of displacement fields are often performed on a cubic grid, and an interpolation step from this grid onto the sheared coordinate system is required. \\

\section{Derivation}

We want to numerically integrate the Takagi-Taupin equations on an orthogonal grid defined by the three orthonormal unit-vectors $\hat{\mathbf{x}}$, $\hat{\mathbf{y}}$, and $\hat{\mathbf{z}}$. The only restriction on the choice of coordinate system is that:

\begin{equation}
    \mathbf{k}_0\cdot \hat{\mathbf{z}} > 0\quad \mathrm{and}\quad \mathbf{k}_h\cdot \hat{\mathbf{z}} > 0
\end{equation}

such that the z-axis takes the role of a quasi-optical axis. We introduce the notation $\mathbf{k}_0 = k_{0,z}\hat{\mathbf{z}} + \mathbf{k}_{0,\perp}$ and similar for $\mathbf{k}_h$ and re-write the TTEs as:

\begin{equation}
 2 k_{0,z}\frac{\partial}{\partial z}E_0(\mathbf{r}) = -ik^2\chi_0E_0(\mathbf{r}) - 2(\mathbf{k}_{0,\perp}\cdot\nabla_\perp)E_0(\mathbf{r})
 -ik^2\chi_{\overline{h}}'(\mathbf{r})E_0(\mathbf{r})
 \label{eq:TT2}
\end{equation}

The equivalent equation for $E_h$ is found by substituting subscripts. We define the transverse-Fourier transform:

\begin{equation}
    \mathcal{F}_\perp\{E(x,y,z)\}(q_x. q_y,z) = \int\int E(x,y,z)\exp(-i2\pi(xq_x + yq_y)) \mathrm{d}x\mathrm{d}y
\end{equation}

With this definition, we Fourier-transform the preceding equation:

\begin{equation}
 \frac{\partial}{\partial z}\tilde{E}_0(q_x,q_y,z) = \Big[ \frac{-ik^2}{2k_{0,z}}\chi_0  - \frac{i2\pi}{k_{0,z}}\mathbf{q}\cdot\mathbf{k}_{0,\perp} \Big] \tilde{E}_0(q_x,q_y,z)  -\frac{ik^2}{2k_{0,z}}{F}_\perp\{\chi_{\overline{h}}'(x,y,z) E_h(x,y,z)\}
\end{equation}

Here we have assumed that $\chi_0$ is constant throughout the simulated volume. We introduce the angles, $\alpha_0$ and $\alpha_h$ given by $\mathbf{k}_0\cdot\hat{\mathbf{z}} = |\mathbf{k}_0|\cos(\alpha_0)$ and similar for \textit{h} to re-write:

\begin{equation}
 \frac{\partial}{\partial z}\tilde{E}_0(q_x,q_y,z) = \Big[ \frac{-ik}{2\cos(\alpha_0)}\chi_0 - \frac{i2\pi}{\cos(\alpha_0)}q_{0,\perp} \Big] \tilde{E}_0(q_x,q_y,z)  -\frac{ik}{2\cos(\alpha_0)}\mathcal{F}_\perp\{\chi_{\overline{h}}'(x,y,z) E_h(x,y,z)\}
    \label{eq:TT3}
\end{equation}

where $q_{0,\perp} = \mathbf{q}\cdot\mathbf{k}_{0,\perp}/k$. \\

In cases where $\chi_h{}'$ is constant or depends only on $z$, the problem can be solved analytically with Green's function methods. In the general case where $\chi_h'$ varies as a function of all  coordinates, the scattering term: $\mathcal{F}_\perp\{\chi_{\overline{h}}'(x,y,z)E_h(x,y,z)\}$ cannot be simplified and we have to use finite difference methods to solve the equations. \\

We note that in cases when both $\mathbf{k}_0$ and $\mathbf{k}_h$ lie within the $x$-$z$ plane, the 2D Fourier transforms may be replaced by 1D Fourier transforms along the x-direction. \\

We introduce a computational grid with axes parallel to the coordinate system. It has step sizes $d_x$, $d_y$ and $d_z$ and number of points, in each dimension, $N_x$, $N_y$ and $N_z$. A point on the grid $P = (i_xd_x, i_yd_y, i_zd_z)$ can be indexes by the numbers $i_x$, $i_y$, $i_z$ where $i_z$ = 0, 1, 2 ... $N_z-1$ and so on. \\

In order to utilize discrete Fourier transform methods when solving these equations on a finite grid, we impose zero Dirichlet boundary conditions in the two transverse dimensions, $x$ and $y$: \\

\begin{equation}
    \begin{split}
        E_0(0,y,z) = E_0(L_x,y,z) = E_h(0,y,z) = E_h(L_x,y,z) &= 0 \\
        E_0(x,0,z) = E_0(x,L_y,z) = E_h(x,0,z) = E_h(x,L_y,z) &= 0 \\
    \end{split}
\end{equation}

\begin{figure}
		\centering
		\includegraphics[scale = 0.5]{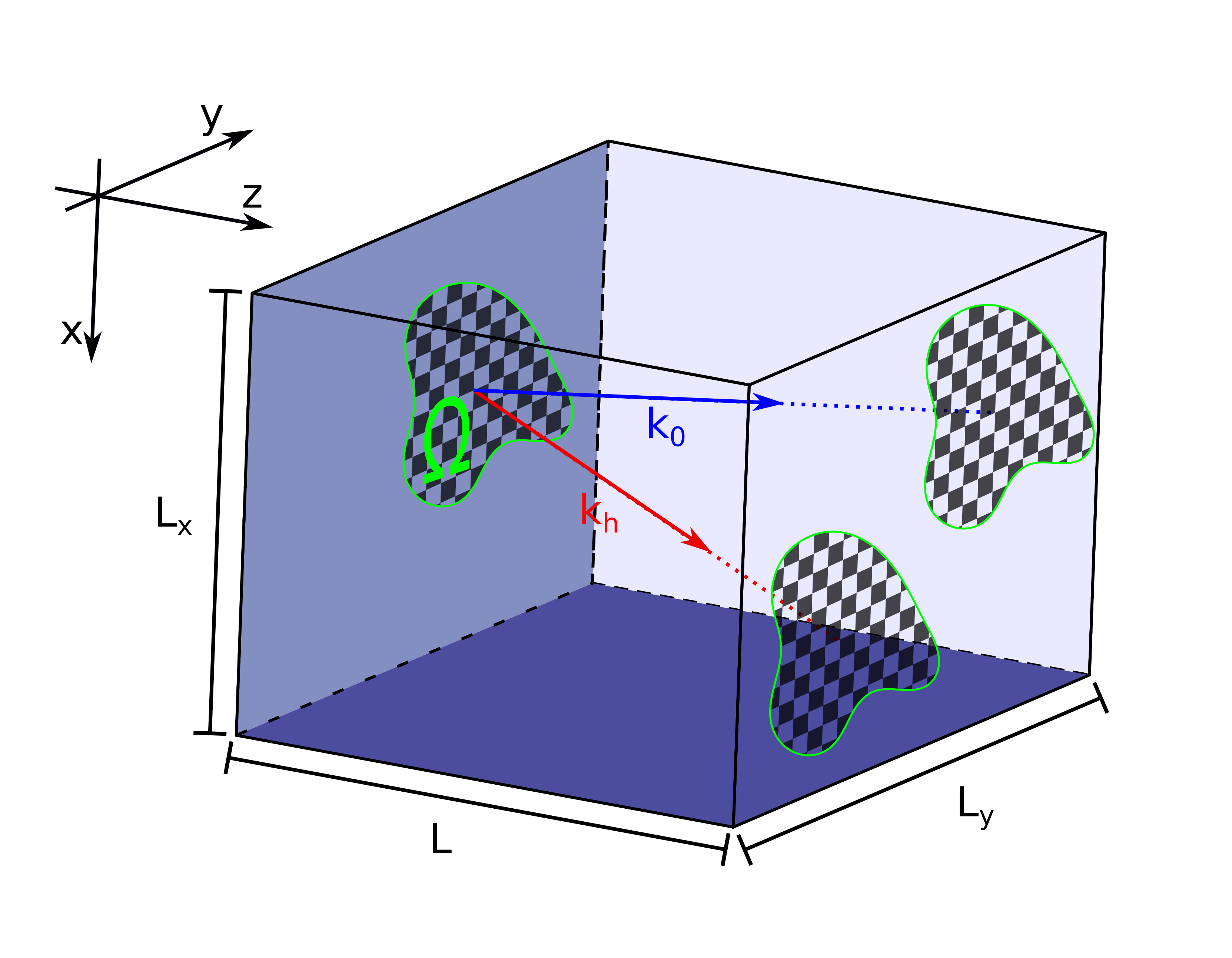}
		\caption{Scattering geometry inside the sample volume and the finite support of the initial condition. }\label{fig:domains}
	\end{figure}

These boundary conditions mean that the sample grid must be large enough to fit the Bormann triangle extending from every point where the initial condition is non-zero. This means that if the initial condition is non-zero only on a domain $\Omega$ on the surface $z=0$, then the direct projection of this domain along the directions of $\mathbf{k}_0$ and $\mathbf{k}_h$ must lie within the sample grid. (see Fig \ref{fig:domains}) \\

	This is fulfilled if the domain $\Omega$ is fully contained in the rectangle defined by:
	
	 \begin{align}
		 \max(0, Lk_{0x}/k, Lk_{hx}/k)< &x < \min(L_x, L_x+Lk_{0x}/k, L_x +Lk_{hx}/k) \\
		 \max(0, Lk_{0x}/k, Lk_{hx}/k)< &y < \min(L_y, L_y+Lk_{0y}/k, L_y+Lk_{hy}/k) 
	 \end{align}
	
which can always be made true for a finitely bounded initial condition if the computational grid is chosen sufficiently large. We use the initial conditions in $z$: $E_h(x,y,0) = 0$ and $E_0(x,y,0) = E_{\mathrm{init}}(x,y)$.	With these boundary conditions, the TTEs become a linear homogeneous initial value problem in $z$ and the integration can be performed with an appropriate finite difference scheme.  \\

\section{Finite difference schemes}

Collecting the discretized components of $\tilde{E}_0(\mathbf{q}_\perp, z)$ and $\tilde{E}_h(\mathbf{q}_\perp, z)$ into a single vector, $\mathbf{E}$, the equations \eqref{eq:TT3} can be written on the form, $\frac{\partial}{\partial z} \mathbf{E} = A \mathbf{E} + \mathbf{B}(z,\mathbf{E})$, where $A$ is a diagonal matrix containing the coefficients in the square brackets of equation \eqref{eq:TT3} and $\mathbf{B}$ contains the convolution terms. In this form, the system of equations can be solved by an exponential integrator, where the $A$-term is handled exactly by an exponential function and the $B$-term is handled by a finite difference scheme.\cite{hochbruck_ostermann_2010} \\

To test the convergence, we utilize two different exponential integrators. The archetypal exponential integrator based on the explicit Euler scheme is given by:
\begin{equation}
\mathbf{E}(z+h) = \exp(hA)\mathbf{E}(z) + h(hA)^{-1}(\exp(hA)-1)\mathbf{B}\left(z,\mathbf{E}(z)\right)
\label{eq:EE}
\end{equation}

Higher order methods can be constructed in a systematic way. One such explicit second order method based on Heun's methods is given by the steps: \cite{MR0494950}
\begin{equation}
\begin{split}
	E^*_1 &= \mathbf{E}(z) \\
	b_1 &= \mathbf{B}(z_n, E_1^*) \\
	E^*_2 &= \phi_0 E_1^* + h\phi_1b_1 \\
	b_2 &= \mathbf{B}(z_n+h, E_2^*) \\
	\mathbf{E}(z+h) &= \phi_0 E_1^* + \frac{h}{2}((2\phi_1-\phi_2)b_1+\phi_2b_2) 
\end{split}
\end{equation}

where the $\phi$-functions are given by: $\phi_0 = \exp(hA)$ and $\phi_n = n(hA)^{-1}(\phi_{n-1}-1)$. \\

The scheme based on Heun's method is chosen here because it only evaluates the $B$ function on the same regular intervals where the field is calculated, and therefore only needs the value of the scattering function on the same grid where the fields are evaluated. \\

For comparison with existing methods we also use a normal finite difference method based on a recent publication by \cite{Shabalin2017} using the half-step finite difference for the derivatives. A derivation of this method is given in Appendix \ref{APP:central_differences}.

\section{Convergence behaviour of exponential methods}
\begin{figure}
	\centering
	\includegraphics[scale = 0.6]{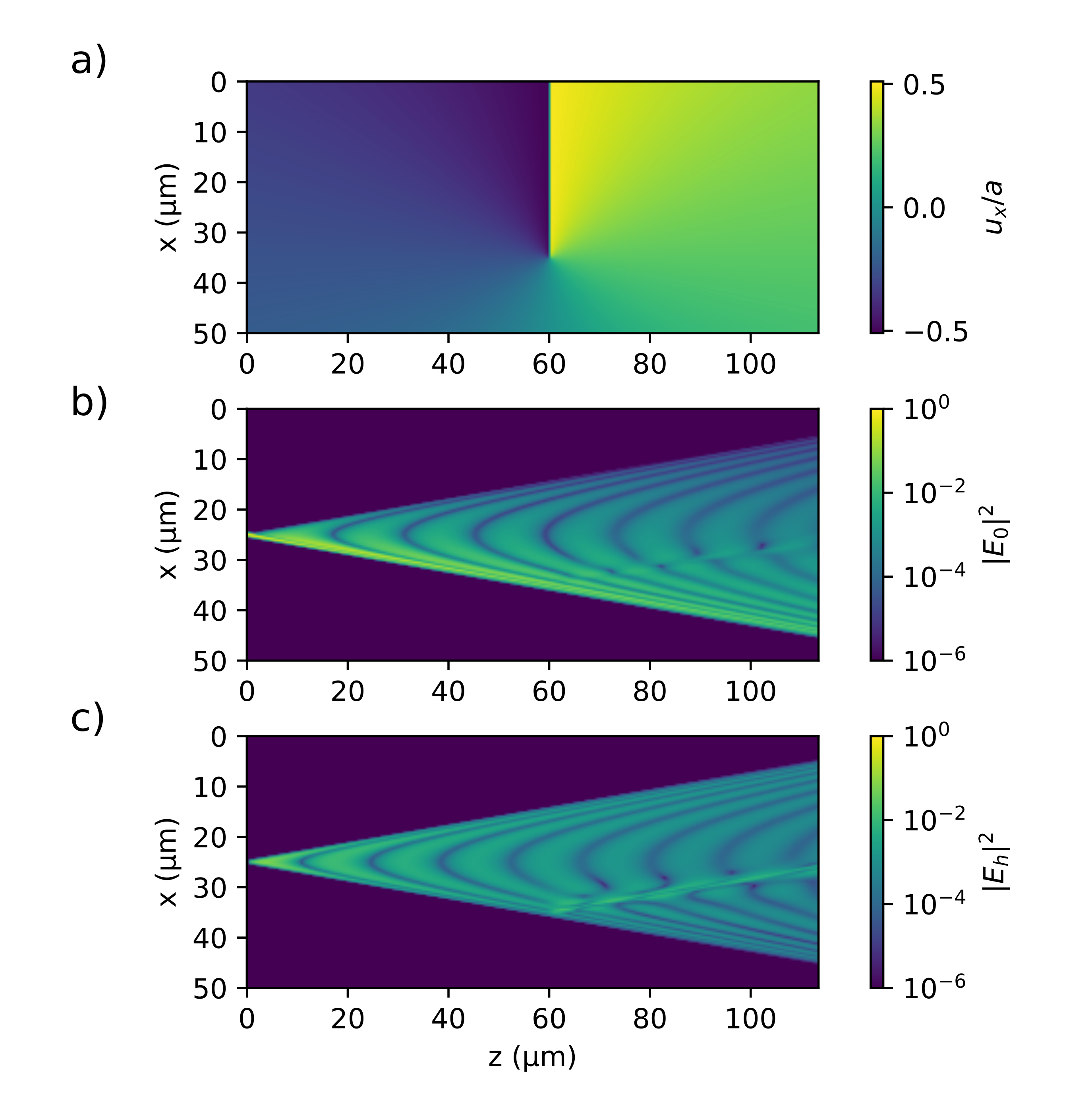}
	\caption{Plots of the sample and calculated fields used in the second convergence test. a) displacement field in units of the lattice constant, $a$. b) Transmitted field on a logarithmic scale, c) scattered field on a logarithmic scale.}
	\label{fig:Direct}
\end{figure}

We choose a symmetric geometry with the scattering vector aligned with the $x$-axis and scattering angle of $2\theta = 21^\circ$, which is typical for hard x-ray diffraction microscopy experiments. We use a sample consisting of a perfect single crystal with a single edge dislocation with Burger's vector (100) close to the path of the direct beam. A plot of the displacement field as well as the amplitudes of the converged solution is shown in figure \ref{fig:Direct}. The fields are simulated under low absorption and highly dynamical conditions. \\

 We simulate only a single slice in the y-direction with the dimensions $50\, \si{\micro m}\times115\, \si{\micro m}$ at a point $1\, \si{\micro m}$ from the dislocation core. The incident beam is a narrow Gaussian of width $\sigma = 0.2\, \si{\micro m}$. \\

In order to accommodate the comparison with existing methods, we utilize a grid with step sizes $\Delta z = h$ and $\Delta x = 2\tan(\theta) h$ for the exponential methods and a grid with the same density of points for the normal finite difference method (see appendix \ref{APP:central_differences}).\\

To check the convergence of the methods we calculate the fields on progressively finer grids. The first grid consisting of only 101$\times$41 points. The error is calculated from the difference on the final slice compared to the solution found with the normal finite difference approach on a very fine grid of 10,241$\times$25,601 steps evaluated on the points of the coarse grid on the exit surface where every grid coincides. \\

\begin{figure}
	\centering
	\includegraphics[scale=0.6]{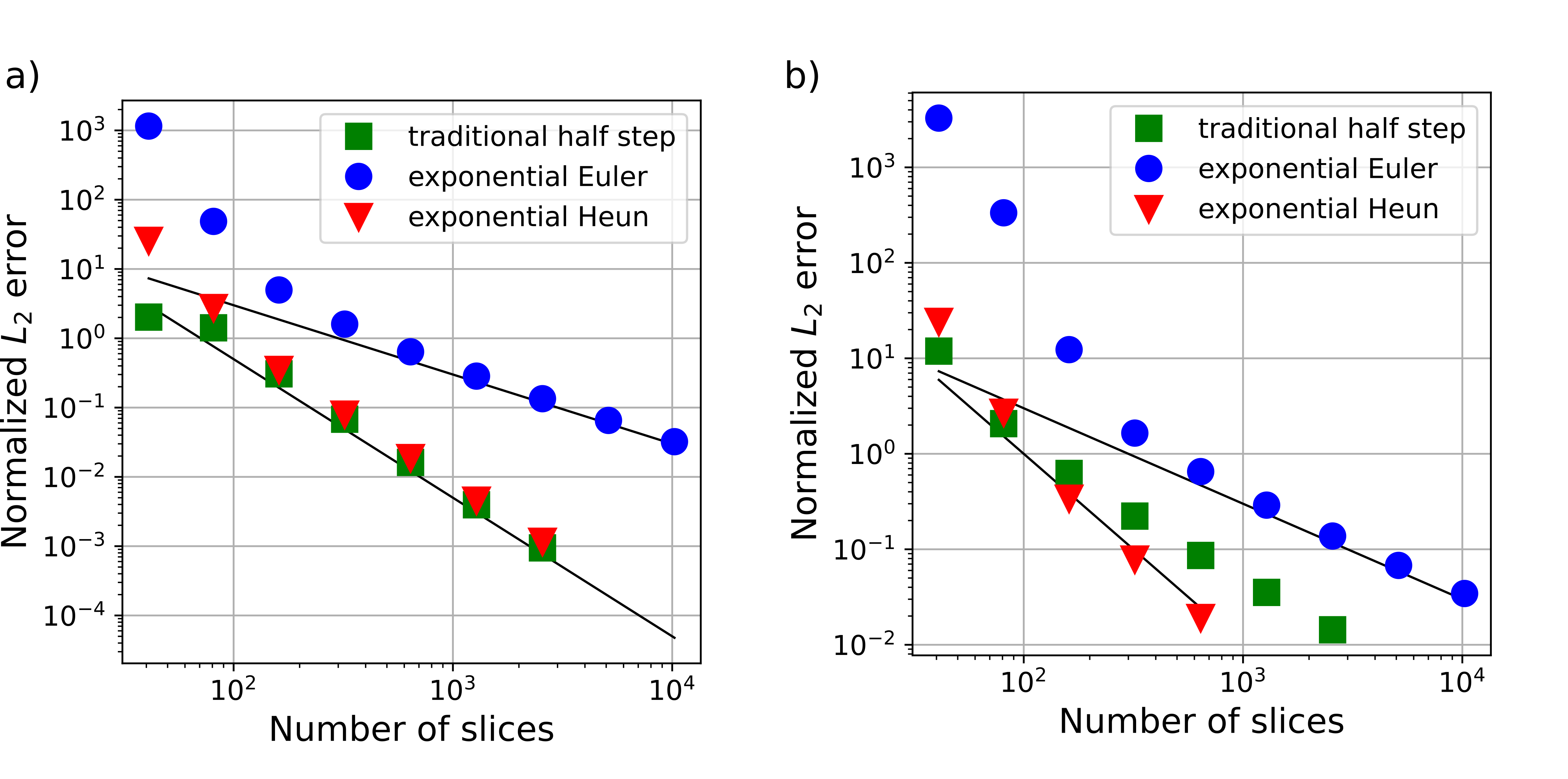}
	\caption{Convergence of the new exponential integrators and a traditional finite difference scheme. The black lines mark first- and second-order convergence respectively. All errors are calculated relative to the solution using the traditional half step method with 10,241 steps. We tested integration schemes on two different samples. One a) is a perfect crystal the other b) is the edge dislocation type sample shown in Fig. \ref{fig:Direct}}
	\label{fig:convergence}
\end{figure}

Figure \ref{fig:convergence} shows the convergence of the three integration methods that all show the expected convergence on a perfect sample a), but the traditional half-step method does not show the expected second order convergence with the edge dislocation sample b). The first order exponential Euler method suffers from an exponential instability and only gives a qualitatively correct result when impractically small step-sizes are utilized. \\

% \section{Including dispersion terms in the TTE's}

% The method as we present it is a mixed real space - reciprocal space method and is conceptually similar to a FFT-based multi slice scheme used to propagate a wave-field through thick optical components. The value of FFT based methods in those situations is that the dispersion of the beam can easily be handled in the paraxial approximation. Typically the dispersion terms are omitted in the TTEs, thus effectively making the projection approximation. \\

% If the dispersion terms are kept, a different version of the TTEs can be written: \\
% \begin{equation}
% \begin{split}
% 2i\mathbf{k}_0\cdot\nabla E_0(\mathbf{r})  & = k^2(\chi_0 E_0(\mathbf{r}) + \chi_{\overline{h}}'(\mathbf{r})E_h(\mathbf{r}) )+ \nabla^2E_0(\mathbf{r})\\
% 2i\mathbf{k}_h\cdot\nabla E_h(\mathbf{r}) & = k^2(\chi_0 E_h(\mathbf{r}) + \chi_{h}'(\mathbf{r})E_0(\mathbf{r}))+ \nabla^2E_h(\mathbf{r}) 
% \end{split}
% \label{eq:TT_disp}
% \end{equation}

% which handles dispersion of the incident and scattered beam inside the crystal. Pair of equations can be solved by the finite difference scheme presented here without any increase of computational complexity. (see Appendix \ref{APP:offaxis}) We find that the approximation made by omitting the dispersion term is small in situations of practical interest but can be significant if the incident radiation contains very high-frequency components.\\

\section{Discussion \& Conclusion}

We have shown a finite difference scheme capable of integrating the TTEs on an orthogonal grid with few restrictions on the choice of grid. We achieve this by implicitly utilizing Fourier interpolation at the level of the individual finite difference step. The method makes approximately the same error as the traditional half-step finite difference scheme. \\

The method utilizes FFTs at each step and has to perform in total 4 2D Fourier transforms (1D in certain geometries) of the entire sample volume, which is expected to be a high computational cost compared to the existing methods. Our experience here, using un-optimized code, is that this increase amounts to about a factor of 4, which unimportant in most cases. \\

The ability to freely choose the computational grid makes implementation of this approach easier, especially when it needs to be combined with other numerical modelling methods. For example if the the input for either the crystal micro structure or the incident field is given by a numerical simulation, or if the scattered fields should be propagated through image-forming optics. \\

\bibliography{biblio}

\newpage
\begin{appendices}
\section{Traditional finite difference scheme}\label{APP:central_differences}

For comparison with the exponential integrators presented in this paper, we also present calculations performed with a 2nd order implicit finite difference scheme based on central difference estimate for the derivatives, which is often applied for dynamical scattering calculations. The derivation here follows the one given by \cite{Shabalin2017} with small changes to the notation.\\

We limit our attention to a symmetric geometry defined by:

\begin{equation}
    \mathbf{k_0} = k\begin{bmatrix}
    \sin\theta\\
    0\\
    \cos\theta\end{bmatrix} \quad \mathrm{and} \quad \mathbf{k_h} = k\begin{bmatrix}
    -\sin\theta \\
    0 \\
    \cos\theta\end{bmatrix}
\end{equation}

Starting from equation \eqref{eq:TT2} we introduce the coordinates:

\begin{equation}
    \begin{bmatrix}
    s_0 \\
    s_h
    \end{bmatrix} = 
    \begin{bmatrix}
    \sin\theta & \cos\theta \\
    -\sin\theta & \cos\theta
    \end{bmatrix}
    \begin{bmatrix}
    x \\
    z
    \end{bmatrix}
\end{equation}

and arrive at a well-known form of the Takagi-Taupin equations: (suppressing the y-dependence)

\begin{equation}
\begin{split}
\frac{\partial E_0(s_0, s_h)}{\partial s_0}  = \frac{k}{2i}\left( \chi_0 E_0(s_0, s_h) + \chi_{\overline{h}}'(s_0, s_h)E_h(s_0, s_h)  \right)  \\
\frac{\partial E_h(s_0, s_h)}{\partial s_h}  = \frac{k}{2i}\left( \chi_0 E_h(s_0, s_h) + \chi_{h}'(s_0, s_h)E_0(s_0, s_h)  \right) 
\end{split}
\label{eq:TT5}
\end{equation}

To avoid truncation errors due the complex rotation caused by the $\chi_0$ terms, we introduce the scaled fields $E_0{}' = \exp\left( \chi_0\frac{ik}{2}s_0 \right)$ and $E_h{}' = \exp\left( \chi_0\frac{ik}{2}s_h \right)$. Plugging in and simplifying some terms gives:

\begin{equation}
\begin{split}
\frac{\partial E_0{}'(s_0, s_h)}{\partial s_0}  = \frac{k}{2i} \exp\left( \chi_0\frac{ik}{2}(s_0 -s_h) \right)\chi_{\overline{h}}'(s_0, s_h)E_h{}'(s_0, s_h)   \\
\frac{\partial E_h{}'(s_0, s_h)}{\partial s_h}  = \frac{k}{2i} \exp\left( \chi_0\frac{ik}{2}(s_h -s_0) \right)\chi_{h}'(s_0, s_h)E_0{}'(s_0, s_h)   
\end{split}
\label{eq:TT6}
\end{equation}

\begin{figure}
	\centering
	\includegraphics[scale=0.7]{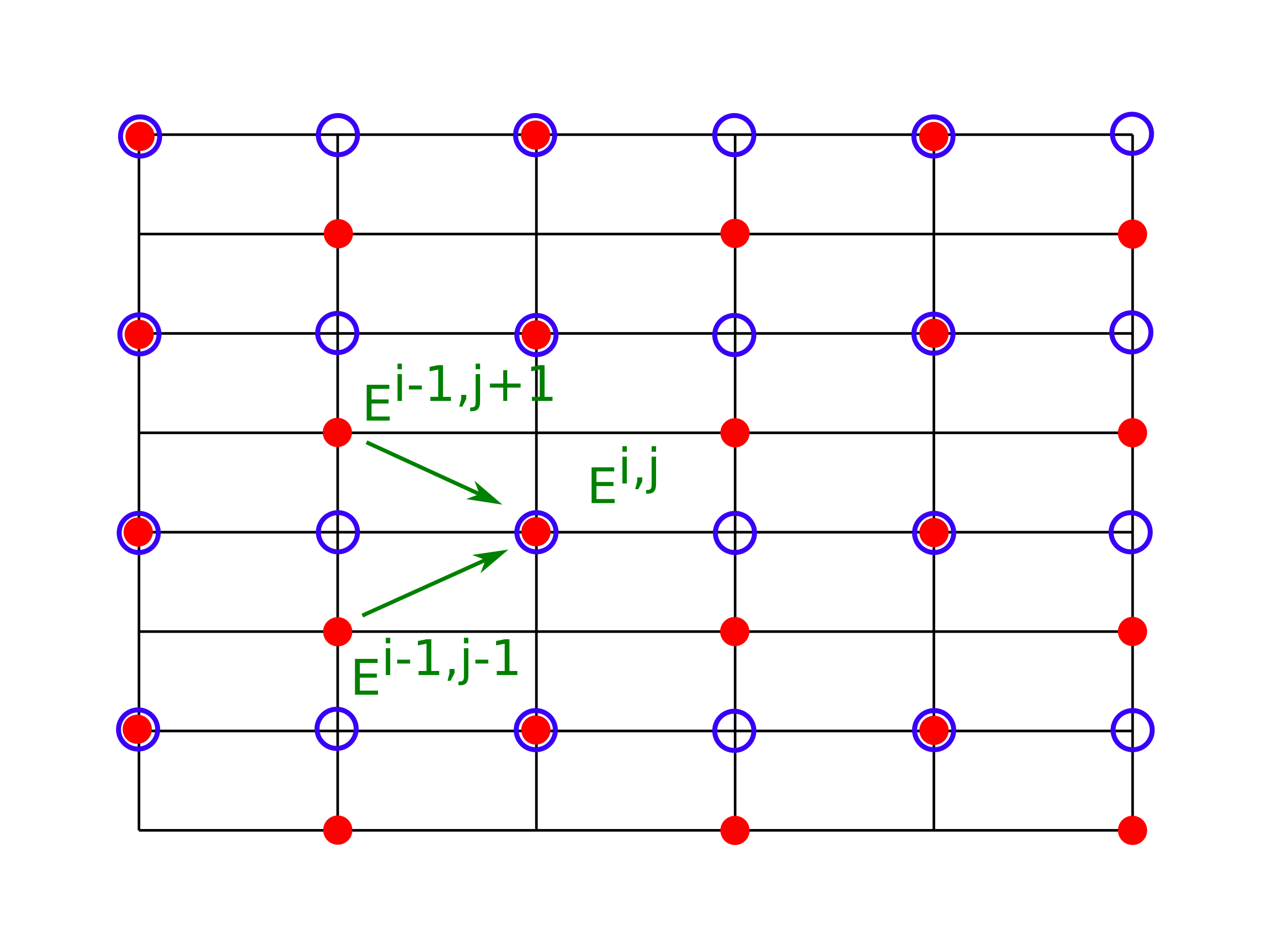}
	\caption{Computational grids used for the traditional half-step approach marked with red dots and for the exponential integrators marked with blue circles. The recurrence relation for the half-step method is drawn with green arrows.}
	\label{fig:recurrence}
\end{figure}

We now introduce a rectangular grid in the original $(x,y)$-coordinates with step-size $h$ in the $z$ direction and $h\tan(\theta)$ in the x-direction. With this choice of grid a subset consisting of every second grid point constitutes a sheared grid aligned with the $s_0$ and $s_h$ directions with both step sizes equal to $p = h(1 + \tan(\theta)^2)^{1/2}$. This allows us to calculate the fields using the finite difference methods and the exponential integrators on grids with the same density of grid points and that coincide on every second plane. We therefore have to choose a grid with an odd number of grid points in the $x$-direction so that we can compare the result on the final slice. (see figure \ref{fig:recurrence})\\

We denote the discretized envelope fields by $E'(x_j, z_i) = E^{i,j}$. The recurrence relation is obtained by the centered first order approximation for the derivatives and a similar centered approximation for the right hand sides in equation \eqref{eq:TT6} to arrive at the equations:

\begin{equation}
\begin{split}
\frac{E_0^{i,j} - E_0^{i-1,j-1}}{p}  = \frac{k}{2i} B\frac{E_h^{i,j} + E_h^{i-1,j-1}}{2}   \\
\frac{E_h^{i,j} - E_h^{i-1,j+1}}{p}  = \frac{k}{2i} D \frac{E_0^{i,j} + E_0^{i-1,j+1}}{2}
\end{split}
\label{eq:TT7}
\end{equation}

where 
\[B = \exp\left( \chi_0\frac{ik}{2}(s^{i,j}_0 -s^{i,j}_h - p/2) \right)\chi_{\overline{h}}\left(s^{i,j}_0-p/2, s^{i,j}_h\right)\] 
and 
\[D = \exp\left( \chi_0\frac{ik}{2}(s^{i,j}_h -s^{i,j}_0 - p/2) \right)\chi_{h}\left(s^{i,j}_0, s^{i,j}_h-p/2\right)\]

introducing the constant $A = \frac{4 i }{kp}$ we can finally write:

\begin{equation}
\begin{split}
E_0^{i,j} = E_0^{i-1,j-1}  + B/A E_h^{i,j} + B/A E_h^{i-1,j-1}   \\
E_h^{i,j} = E_h^{i-1,j+1}  + D/A E_0^{i,j}+ D/A  E_0^{i-1,j+1}
\end{split}
\label{eq:recurrence}
\end{equation}

which are the implicit recurrence relations used in the calculations. Furthermore we need the boundary conditions, that the fields are both zero at the top and bottom surfaces: $E_{0/h}^{i,0} = E_{0/h}^{i,N_x-1} = 0$

\end{appendices}

\end{document}